\documentclass[final,12pt,a4paper]{elsart}

\usepackage{amsmath,amssymb} 
\usepackage[dvips]{graphicx} 

\def\epsilon{\varepsilon}
\def\phi{\varphi}

\begin{document}

\begin{frontmatter}


\title{Preparation of Schr\"{o}dinger cat states with cold ions in a cavity beyond the
Lamb-Dicke limit}

\author[dfreitas]{Dagoberto S. Freitas}
\ead{dfreitas@uefs.br} \ead[url]{http://www.uefs.br}
\author[jrocha]{Jairo R. de Oliveira}
\ead{jrocha@df.ufrpe.br} \ead[url]{http://www.ufrpe.br}
\address[dfreitas]{Departamento de Fisica, Universidade Estadual de Feira de Santana,
44036-900 Feira de Santana, BA, Brazil}
\address[jrocha]{Departamento de Fisica, Universidade Federal Rural de Pernambuco,
52171-900 Recife, PE, Brazil}

\begin{abstract}
We investigate the dynamics of a cold trapped ion coupled to the
quantized field inside a high-finesse cavity. We have used an
approach for preparing the {\bf SC} states of motion of ion. This
approach, based on unitary transformating the Hamiltonian, allows
its exact diagonalization without performing the Lamb-Dicke
aproximation. We show that is possible to generate a {\bf SC} states
having rather simple initial state preparation, e.g., the vacuum
sate for both cavity field and the ion motion.
\end{abstract}

\end{frontmatter}


The manipulation of simple quantum systems such as trapped ions
\cite{wine98} has opened new possibilities regarding not only the
investigation of foundations of quantum mechanics, but also
applications on quantum information. In such a system, the internal
degrees of freedom of an atomic ion may be coupled to the
electromagnetic field as well as to the motional degrees of freedom
of the ion's center of mass. This system allows the preparation of
nonclassical states of the vibrational motion of the ion
\cite{cira1,vogel}. In fact, the generation of Fock, coherent,
squeezed, \cite{meek} and Schr\"{o}dinger-cat states \cite{monr} has
been already accomplished. Under certain circumstances, in which
full quantization of the three sub-systems becomes necessary, is
reasonable to assume that the single trapped ion interacting with a
quantized cavity field in such a way that the quantized field gets
coupled to the atom. The quantization of the field of course brings
new possibilities. Within that realm, it has been already
investigated the influence of the field statistics on the ion
dynamics \cite{zeng94,knight98}, as well as the transfer of
coherence between the motional states and the field
\cite{parkins99}, a scheme for generation of matter-field Bell-type
states \cite{ours01}, and even propositions of quantum logic gates
\cite{gates}.

In this paper we explore further the consequences of having the
trapped ion in interaction with a quantized field. Here we adopt an
approach to this problem. We depart from the full ion-field cavity
Hamiltonian, and perform a unitary transformation that allows us to
obtain a diagonalization Hamiltonian without the rotating-wave
approximation (RWA)\cite{ours99}. We try to prepare the {\bf SC}
states beyond the LDL. The {\bf SC} state, i.e., the superposition
of macroscopically distinguishable states. Experimentally,
\textbf{SC} states have been obtained in NIST group with single cold
$^{9}Be^{+}$\cite{gerry97}.


We consider a single trapped ion, within a Paul trap, placed inside
a high finesse cavity, and having a cavity mode coupled to the
atomic ion \cite{semiao02}. The vibrational motion is also coupled
to the field as well as to the ionic internal degrees of freedom, in
such a way that the hamiltonian will reads \cite{knight98,ours01}.
\begin{equation}
\hat{H}=\hbar\nu \hat{a}^{\dagger}\hat{a} +
\hbar\omega\hat{b}^{\dagger}\hat{b} +\hbar\frac{\omega_0}{2}\sigma_z
+\hbar g(\sigma_+ + \sigma_-)(\hat{b}^{\dagger}+
\hat{b})\cos\eta(\hat{a}^{\dagger}+\hat{a}), \label{H}
\end{equation}
where $\hat{a}^{\dagger}(\hat{a})$ denote the creation
(annihilation) operators of the center-of-mass vibrational motion of
the ion (frequency $\nu$), $\hat{b}^{\dagger}(\hat{b})$ are the
creation (annihilation) operators of photons in the field mode
(frequency $\omega$), $\omega_0$ is the atomic frequency transition,
$g$ is the ion-field coupling constant, and $\eta=2\pi a_0/\lambda$
is the Lamb-Dicke parameter, being $a_0$ the amplitude of the
harmonic motion and $\lambda$ the wavelength of light. Typically the
ion is well localized, confined in a region much smaller than
light's wavelenght, or $\eta\ll 1$ (Lamb-Dicke regime). Usually
expansions up to the first order in $\eta$ are made in order
simplify hamiltonians involving trapped ions, which results in
Jaynes-Cummings like hamiltonians. However, even for small values of
the Lamb-Dicke parameter, an expansion up to second order in $\eta$
interesting effects, such as long time scale revivals, are
observed\cite{semiao02}.

We first applying the unitary transformation\cite{ours99}
\begin{equation}
\hat{T} = {1\over \sqrt{2}} \left\{ {1\over 2} \left[
\hat{D}^{\dagger}\left(\beta \right) + \hat{D}\left(\beta \right)
\right]\hat{I} + {1\over 2} \left[
\hat{D}^{\dagger}\left(\beta\right) -\hat{D}\left(\beta \right)
\right] \hat{\sigma}_z +\hat{D}\left(\beta \right)\hat{\sigma}_{+}
-\hat{D}^{\dagger}\left(\beta \right) \hat{\sigma}_{-}
\right\},\label{T}
\end{equation}
to the Hamiltonian in Eq. (\ref{H}), where
$\hat{D}(\beta)=\exp(\beta\hat{a}^\dagger- \beta^*\hat{a})$ is
Glauber's displacement operator, with $\beta=i\eta/2$, we obtain the
following transformed Hamiltonian
\begin{equation}
\hat{\cal H} \equiv \hat{T}\hat{H}\hat{T}^{\dagger} = \hbar \nu
\hat{a}^{\dagger}\hat{a}+ \hbar \omega \hat{b}^{\dagger}\hat{b} +
\hbar g(\hat{b}^{\dagger}+\hat{b})\hat{\sigma}_z
-i\hbar{\eta\nu\over 2} \left[ (\hat{a}^{\dagger}-\hat{a})
-i\frac{\omega_o}{\eta\nu}\right]
\left(\hat{\sigma}_{-}+\hat{\sigma}_{+}\right)+ \hbar \nu {\eta^2
\over 4}.\label{HT0}
\end{equation}
This result holds for any value of the Lamb-Dicke parameter $\eta$.
In the regime $\eta\nu\gg g$, i.e., when the ion-field coupling
constant is much smaller than the frequency of the ion in the trap
($\nu$). In this regime the Hamiltonian in Eq.(\ref{HT0})will become
\begin{equation}
\hat{\cal H} = \hbar \nu \hat{a}^{\dagger}\hat{a}+ \hbar \omega
\hat{b}^{\dagger}\hat{b} -i\hbar{\eta\nu\over 2} \left[
(\hat{a}^{\dagger}-\hat{a}) -i\frac{\omega_o}{\eta\nu}\right]
\left(\hat{\sigma}_{-}+\hat{\sigma}_{+}\right)+ \hbar \nu {\eta^2
\over 4}.\label{HT1}
\end{equation}
The time evolution of the state vector, for an initial state
$|\psi(0)\rangle$ is
\begin{eqnarray}
|\psi(t) \rangle & = & \hat{T}^\dagger\hat{U}_{T}(t)\hat{T}|\psi(0)\rangle \nonumber
\\
\\
& = & \hat{U}(t)|\psi(0)\rangle ,\nonumber
\end{eqnarray}
where $\hat{U}_{T}(t)=\exp{(-i\hat{\cal H}t/\hbar)}$ is the
evolution operator in the transformation representation and
$\hat{U}(t)= \hat{T}^\dagger\hat{U}_{T}(t)\hat{T}$ is the time
evolution operator in the original representation. After some
algebra, the time evolution operation in the original representation
is
\begin{eqnarray}
\hat{U}(t)&=& \frac{e^{-i\omega\hat{b}^\dagger \hat{b}t}
e^{-i\nu\hat{a}^\dagger \hat{a}t}}{2} \left\{
\left[D^\dagger(\beta)+D(\beta)\right]-
\left[D^\dagger(\beta)-D(\beta)\right]\sigma_{z}\right\} \nonumber \\
&&\left[\cos\left(\frac{\omega_{o}t}{2}\right)+
i\sigma_{x}\sin\left(\frac{\omega_{o}t}{2}\right)\right]
\end{eqnarray}
Consider that the state vector having the following initial
condition for the ion-field state
\begin{equation}
|\psi(0) \rangle=|0\rangle_{\nu}|0\rangle_{\omega}\left[{1\over
\sqrt{2}}\left(|e\rangle + |g\rangle \right)\right],
\end{equation}
or the field prepared in a vacuum state $|0\rangle_{\omega}$ and the
ion's center-of-mass motion prepared in a vacuum state
$|0\rangle_{\nu}$, and the ion's internal levels prepared in a
superposition of atomics states. The time evolution of the state
vector is given by
\begin{equation}
|\psi(t) \rangle = \frac{e^{-\frac{i\omega_{o}t}{2}}}{\sqrt{2}}
\left[|e^{-i\nu t}\beta\rangle|e\rangle + |-e^{-i\nu
t}\beta\rangle|g\rangle\right]|0\rangle_{\omega}.
\end{equation}
Applying a pulse\cite{feng01}
\begin{equation}
 \hat{V} = {1\over \sqrt{2}}\left(
                              \begin{array}{cc}
                                1 & 1 \\
                                -1 & 1 \\
                              \end{array}
                            \right)
\end{equation}
on the ion the superposition of coherent states will be obtained
 \begin{equation}
\hat{V}|\psi(t) \rangle =
\frac{e^{-\frac{i\omega_{o}t}{2}}}{\sqrt{2}} \left(\Phi_{+}|e\rangle
+ \Phi_{-}|g\rangle\right)|0\rangle_{\omega}
 \end{equation}
with the {\bf SC}
\begin{equation}
\Phi_{\pm} = {1\over \sqrt{2}}\left(|e^{-i\nu t}\beta\rangle \pm
|-e^{-i\nu t}\beta\rangle\right).
\end{equation}
The resulting state above is an entangled state involving the ion's
internal (electronic) degrees of freedom, the vibrational motion and
the cavity field. If one measures the internal state of the ion
(either in $|g\rangle$ or $|e\rangle$) and the cavity field in the
vacuum state, that action will collapse the $\hat{V}|\psi(t)
\rangle$ in a {\bf SC} states $\Phi_{\pm}$.

In summary, we have presented an approach for preparing the {\bf SC}
states of motion of cold trapped ion placed inside a high-Q cavity.
This approach, based on unitary transformating the Hamiltonian,
allows its exact diagonalization without performing the Lamb-Dicke
aproximation. We have shown that it is possible to generate a {\bf
SC} states having rather simple initial state preparation, e.g., the
vacuum sate for both cavity field and the ion motion.
%
%

%
%

\end{document}